\documentclass[letter]{aa} 

\usepackage{graphicx}
\usepackage{hyperref}
\hypersetup{
    colorlinks=true,
    citecolor=blue,
    linkcolor=blue,
    filecolor=magenta,      
    urlcolor=blue,
}

\usepackage{txfonts}
\DeclareUnicodeCharacter{2212}{-}
\usepackage{natbib}
\bibpunct{(}{)}{;}{a}{}{,}

\def\kms  {km\,s$^{-1}$}

\def\bco {\ifmmode{^{12}{\rm CO}(J=2\to1)}\else{$^{12}{\rm CO}(J=2\to1)$}\fi}

\def\mh     {H$_{2}$}

\def\Lsun{L$_{\odot}$}
\def\Linelum{L$_{\odot}$~pc$^{-2}$}
\def\Msun{M$_{\odot}$}
\def\deg {$^{\circ}$}

\def\galaxy{ESO~320-G030}

\begin{document}

\title{ A spectacular galactic scale magnetohydrodynamic powered wind in ESO~320-G030}

   \author{M. D. Gorski\inst{\ref{inst.OSO},\ref{inst.CIERA}}
          \and
          S. Aalto\inst{\ref{inst.OSO}}
          \and
          S. K{\"o}nig\inst{\ref{inst.OSO}}
          \and
          C. F. Wethers\inst{\ref{inst.OSO}}
          \and
          C. Yang\inst{\ref{inst.OSO}}
          \and
          S. Muller\inst{\ref{inst.OSO}}
          \and
          K. Onishi\inst{\ref{inst.OSO},\ref{inst.NAOJ}}
          \and
          M. Sato\inst{\ref{inst.OSO}}
          \and
          N. Falstad\inst{\ref{inst.OSO}}
          \and
          Jeffrey G.~Mangum\inst{\ref{inst.NRAOcvill}}
          \and
          S. T. Linden\inst{\ref{inst.UMASS}}
          \and
          F. Combes\inst{\ref{inst.ObsParis}}
          \and
          S. Mart\'in \inst{\ref{inst.ESOChile},\ref{inst.JAO}}
          \and 
          M. Imanishi\inst{\ref{inst.NAOJ}}         
          \and
          Keiichi Wada\inst{\ref{inst.Kagoshima}}
          \and
          L. Barcos-Mu\~noz\inst{\ref{inst.NRAOcvill},\ref{inst.UVa}}
          \and
          F. Stanley\inst{\ref{inst.IRAM}}
          \and
          S. Garc\'{\i}a-Burillo\inst{\ref{inst.OANIGN}}
          \and
          P.P.\ van der Werf\inst{\ref{inst.Leiden}}
          \and
          A. S. Evans\inst{\ref{inst.NRAOcvill},\ref{inst.UVa}}
          \and
          C. Henkel \inst{\ref{inst.MPIfR},\ref{inst.Abdulaziz},\ref{inst.Xinjiang}}
          \and
          S. Viti\inst{\ref{inst.Leiden},\ref{inst.UCL}}
          \and
          N. Harada \inst{\ref{inst.NAOJ},\ref{inst.ASP}}
          \and
          T. D\'{\i}az-Santos \inst{\ref{inst.FORTH},\ref{inst.SOSC}}
          \and 
          J. S. Gallagher \inst{\ref{inst.WISC}}
          \and
          E. Gonz\'alez-Alfonso \inst{\ref{inst.alcala}}
          }

   \institute{
            \label{inst.OSO}Department of Space, Earth and Environment, 
            Chalmers University of Technology, Onsala Space Observatory, 
            439 92 Onsala, Sweden \\
            \email{mark.gorski@northwestern.edu}
            \and
            \label{inst.NRAOcvill}National Radio Astronomy Observatory, 520 Edgemont Road,
            Charlottesville, VA  22903-2475, USA
            \and
            \label{inst.UMASS}Department of Astronomy, University of Massachusetts at Amherst, Amherst, MA 01003, USA
            \and
            \label{inst.ObsParis}Observatoire de Paris, LERMA, Collège de France, PSL University, CNRS, Sorbonne University, F-75014 Paris, France
            \and
            \label{inst.ESOChile}European Southern Observatory, Alonso de C\'ordova, 3107, Vitacura, Santiago 763-0355, Chile
            \and
            \label{inst.JAO}Joint ALMA Observatory, Alonso de C\'ordova, 3107, Vitacura, Santiago 763-0355, Chile
            \and
            \label{inst.NAOJ}National Astronomical Observatory of Japan, 2-21-1, Osawa, Mitaka, Tokyo, 181-8588, Japan 
            \and
            \label{inst.Kagoshima}Kagoshima University, 1-21-35, Kagoshima, 890-0065, Japan
            \and
            \label{inst.UVa}Department of Astronomy, University of Virginia, 530 McCormick Road, Charlottesville, VA 22903, USA
            \and
            \label{inst.IRAM}Institut de Radioastronomie Millim\'etrique (IRAM), 300 Rue de la Piscine, 38400 Saint-Martin-d'H\`eres, France
            \and
            \label{inst.OANIGN}Observatorio Astron\'omico Nacional (OAN-IGN), Observatorio de Madrid, Alfonso XII, 3, 28014, Madrid, Spain
            \and
            \label{inst.Leiden}Leiden Observatory, Leiden University, PO Box 9513, 2300 RA Leiden, The Netherlands
            \and
            \label{inst.MPIfR}Max-Planck-Institut f\"ur Radioastronomie, Auf-dem-H\"ugel 69, 53121 Bonn, Germany
            \and
            \label{inst.Xinjiang}Xinjiang Astronomical Observatory, Chinese Academy of Sciences, 830011 Urumqi, China
            \and
            \label{inst.Abdulaziz}Astron. Dept., Faculty of Science, King Abdulaziz University, P.O. Box 80203, Jeddah 21589, Saudi Arabia
            \and
            \label{inst.UCL}Department of Physics and Astronomy, University College London, Gower Street, London, WC1E 6BT, UK
            \and
            \label{inst.ASP}Astronomical Science Program, Graduate Institute for Advanced Studies, SOKENDAI, 2-21-1 Osawa, Mitaka, Tokyo 181-1855, Japan
            \and
            \label{inst.FORTH}Institute of Astrophysics, Foundation for Research and Technology-Hellas (FORTH), Heraklion, 70013, Greece
            \and
            \label{inst.SOSC}School of Sciences, European University Cyprus, Diogenes street, Engomi, 1516 Nicosia, Cyprus
            \and
            \label{inst.WISC}Department of Astronomy, University of Wisconsin, 475 North Charter St., Madison, WI 53706 USA
            \and
            \label{inst.alcala} Universidad de Alcal\'a, Departamento de F\'{\i}sica y Matem\'aticas, Campus Universitario, E-28871 Alcal\'a de Henares, Madrid, Spain 
            \and
            \label{inst.CIERA}Center for Interdisciplinary Exploration and Research in Astrophysics (CIERA) and Department of Physics and Astronomy, Northwestern University, Evanston, IL 60208, USA
            }

   \date{recived \today }


\abstract{
How galaxies regulate nuclear growth through gas accretion by supermassive black holes (SMBHs) is one of the most fundamental questions in galaxy evolution.
One potential way to regulate nuclear growth is through a galactic wind that removes gas from the nucleus.
It is unclear whether galactic winds are powered by jets, mechanical winds, radiation, or via magnetohydrodynamic (MHD) processes.
Compact obscured nuclei (CONs) represent a significant phase of galactic nuclear growth.
These galaxies hide growing SMBHs or unusual starbursts in their very opaque, extremely compact (r $<$ 100 pc) centres.
They are found in approximately 30\% of the luminous and ultra-luminous infrared galaxy (LIRG and ULIRG) population.
Here, we present high-resolution ALMA observations ($\sim$30~mas, $\sim$5~pc) of ground-state and vibrationally excited HCN towards \galaxy\ (IRAS 11506−3851). 
\galaxy\ is an isolated luminous infrared galaxy known to host a compact obscured nucleus and a kiloparsec-scale molecular wind. 
Our analysis of these high-resolution observations excludes the possibility of a starburst-driven wind, a mechanically or energy driven active galactic nucleus (AGN) wind, and exposes a molecular MDH wind. 
These results imply that the nuclear evolution of galaxies and the growth of SMBHs are similar to the growth of hot cores or protostars where gravitational collapse of the nuclear torus drives a MHD wind.
These results mean galaxies are capable, in part, of regulating the evolution of their nuclei without feedback.
}

  
\maketitle

\section{Introduction \label{sec:intro}}
A fundamental process in galaxy evolution is supermassive black hole (SMBH) growth \citep[e.g.,][]{Sanders1996, Ferrarese2000, Fabian2012}.
One potential way to regulate nuclear growth is through feedback.
Energy and momentum injected into the interstellar medium (ISM) from intense episodes of star formation \citep[e.g.,][]{Lehnert1999, Leroy2018, Bolatto2021} and active galactic nuclei~(AGNs) \citep[e.g.,][]{Garcia-Burillo2014} can hinder nuclear growth by removing gas in the form of galactic winds, which are commonly observed in galaxies \citep{Veilleux2020}.
Astronomers currently debate whether AGN-driven galactic outflows are fuelled by jets, mechanical winds, or radiation \citep{Faucher-Giguere2012, Wada2012, Costa2014, Veilleux2020, Ishibashi2021}.
Magnetohydrodynamic (MHD) winds, which are not the result of feedback, could also regulate nuclear growth \citep[e.g.,][]{Chan2017,Vollmer2018,Girichidis2018, Takasao2022}, as suggested by \citet{aalto2020} within the galaxy NGC 1377.
MHD winds remove material from the nucleus, but also allow for accretion onto a central object by removing angular momentum from the system \citep{Blandford1982,Ray2021}.
Although radio jets are a MHD phenomenon that may entrain molecular gas \citep[e.g.,][]{Morganti2007},  molecular MHD winds are a less widely accepted explanation for the mechanism behind galactic winds.

Hiding either growing SMBHs or unusual starbursts, compact obscured nuclei (CONs) represent a key phase of galactic nuclear evolution \citep{Aalto2015b}.
The most intense phase of SMBH growth is theorized to occur when it is profoundly dust enshrouded \citep[][\& references therein]{Hickox2018}. 
CONs are galaxy nuclei that are compact (r < 100 pc), hot (T > 100 K), and opaque (N(H$_2$) $>$ 10$^{24}$~cm$^{-2}$), and identified by a luminosity surface density of the vibrationally excited HCN~$v2=1f$ $J=3\rm{-}2$ (HCN-vib) transition (rest frequency 265.85270940~GHz) greater than  1~\Lsun~pc$^{-2}$.
It is established that about $\sim$40\% of ultra-luminous infrared galaxies (ULIRGs) and $\sim$20\% of luminous infrared galaxies (LIRGs) contain CONs within the nearby universe \citep[$D_L \lesssim 100$ Mpc,][]{Falstad2021}.
CON nuclei are thought to be almost completely opaque with extinctions of $\rm{A_v}>1000$ \citep[e.g.,][]{Treister2010,Roche2015}.
These galaxies provide an opportunity for us to investigate the correlations between the nuclear growth processes in galaxies, SMBHs, and global galaxy characteristics.

\citet{Falstad2021} identified \galaxy\ (IRAS F11506-3851) as a host to a CON with a HCN-vib luminosity of 1.4$\pm$0.2~\Linelum.
\galaxy\ is an isolated LIRG \citep{Sanders2003}, with regular rotation \citep{Bellocchi2013}, a double-barred structure \citep{Greusard2000}, and no signs of a recent interaction \citep{Arribas2008}.
It has a luminosity distance of 36~Mpc ($z = 0.0103$) \citep{Sanders2003} using a flat cosmology with H$_0 = 75$~\kms~Mpc$^{-1}$, $\Omega_{\rm{M}} = 0.3$, and $\Omega_\Lambda = 0.7$.
The systemic velocity is $\sim3080$~\kms~ \citep{Pereira-Santaella2016} and it has an infrared luminosity of $\sim10^{11}$~\Lsun~ \citep{Sanders2003}.
It lacks traditional indications of an AGN based on X-ray or mid-infrared observations \citep{Pereira-Santaella2010b,Pereira-Santaella2011}, and it lacks the flat radio spectrum of a radio-loud AGN \citep{Baan2006} to power its OH megamaser \citep{Staveley-Smith1992}.
The galaxy harbours a multiphase outflow observed from 100~pc to kiloparsec scales \citep[e.g.,][]{Arribas2014, Cazzoli2014, Pereira-Santaella2016, Pereira-Santaella2020}, but the source of this outflow is claimed to be within the innermost 250~pc of the galaxy's nucleus \citep{Pereira-Santaella2020}.

In this work we present high-resolution ($\sim$30~mas, $\sim$5~pc) Atacama Large Millimeter/Submillimeter Array (ALMA) observations of \galaxy. 
Due to the proximity of \galaxy, and the high quality of the ALMA data, we are able to reveal the launching region of the molecular outflow. 
We present the properties of the observations in Section \ref{sec:observations}, and the observed properties of both ground-state and vibrationally excited HCN transitions as well as the continuum emission in Section \ref{sec:results}. 
In Section \ref{sec:discussion} we discuss the nuclear -- and outflow -- properties of \galaxy.
Our findings on the wind origin are summarized in Section \ref{sec:conclusions}.

  
\section{Observations \label{sec:observations}}

Our ALMA observations are part of the CONfirm project (PI Falstad, N., project code 2019.1.01612.S) which contain the HCN \,$\nu$=0, $J$=$3\rm{-}2$ (rest frequency 265.8864343~GHz) and HCN~$\nu2=1f$, $J$=$3\rm{-}2$ transitions (hereafter HCN-vib, rest frequency 267.19928300~GHz) towards \galaxy.
We combine three scheduling blocks, two from September 6, 2021, and  one from September 28, 2021.
The three scheduling blocks utilized 42, 42, and 44 antennas, respectively.
The respective average precipitable water vapour was $\sim2.1$~mm, $\sim2.2$~mm, and $\sim0.7$~mm.
The longest baseline was 16.2~km corresponding to the highest possible angular resolution of 20~mas at 260~GHz.
The fifth percentile baseline length corresponds to a maximum recoverable scale of 0\farcs4.

The calibration of and imaging of the data were done in CASA \citep{CASATeam2022PASP}.
The spectral setup consists of two 1.875~GHz spectral windows containing 240 channels centred at 265.005~GHz and 263.182~GHz.
Two additional 2.000~GHz spectral windows with 128 channels are centred at 248.447~GHz and 250.427~GHz.
The original synthesized beam is 30.9~mas~$\times$~28.8~mas with a natural weighting of the visibilities. 
The final image cubes have 18~MHz wide ($\sim 20$~\kms) channels imaged to a common beam of 31~mas~$\times$~31~mas (5.3~pc~$\times$~5.3~pc at a distance of 36~Mpc) with 6.5~mas pixels, resulting in a root-mean-square (rms) of 0.22 mJy beam$^{-1}$.

\section{Results \label{sec:results}}

\subsection{Continuum}
\begin{figure}[]
    \centering
    \includegraphics[width=0.49\textwidth]{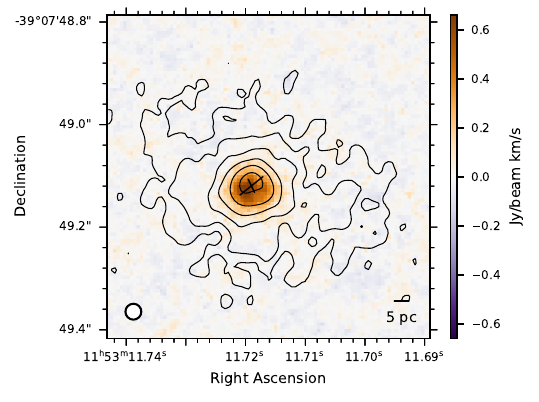}
    \caption{ Colour image showing the integrated flux map of the vibrationally excited HCN~$\nu2=1f$, $J$=$3\rm{-}2$ transition. 
    The contours show the 256~GHz continuum with levels at 0.02~$\times$ (3, 6, 12, 24, 48, and 96)~mJy~beam$^{-1}$.
    The cross indicates the position and orientation of the  68.2~mas by 30.5~mas Gaussian fit to the continuum peak.
    }
    \label{fig:Continuum}
\end{figure}

We have combined the two 2.0~GHz spectral windows and the line-free channels from the 1.875~GHz spectral windows to generate a 256~GHz continuum image (Figure \ref{fig:Continuum}) with a rms of 21~$\mu$Jy~beam$^{-1}$. 
A two-dimensional Gaussian fit to the continuum emission, masking pixels below 80\% of the peak intensity, reveals a central deconvolved source with a full width at half maximum (FWHM) of 68.2$\pm$0.8~mas by 30.5$\pm$0.4~mas at a position angle of 120.8$\pm0.5$\deg.
At a distance of 36~Mpc, the physical size is 11.9$\times$5.3~pc, that is, of a similar size to nearby dusty AGN tori \citep{Combes2019,Garcia-burillo2021}.
\subsection{ HCN emission}

We used the {\tt STATCONT} software package \citep{SanchezMonge2018} to perform the continuum subtraction on the image cubes.
STATCONT statistically derives the continuum on a pixel-by-pixel basis.
We utilized the sigma clipping method and specify a rms of 0.22 mJy beam$^{-1}$ per channel.

\begin{figure}[]
    \centering
    \includegraphics[width=0.49\textwidth]{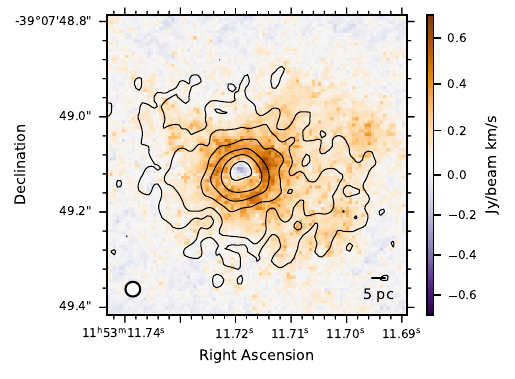}
    \caption{Integrated flux map (moment 0) of the ground-state  HCN~$J=3\rm{-}2$ transition. 
    The contour levels are 0.039~$\times$ (4, 8, 16, and 32) Jy~beam$^{-1}$~\kms.
    }
    \label{fig:HCN_mom0}
\end{figure}

\begin{figure}[]
    \centering
    \includegraphics[width=0.49\textwidth]{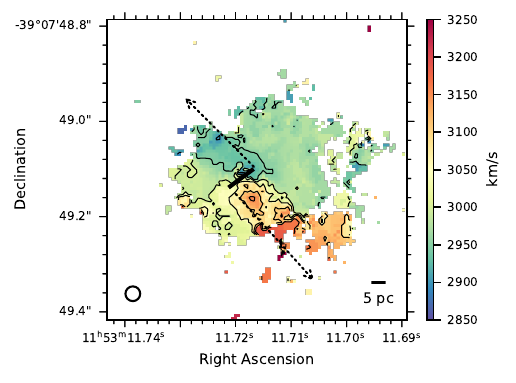}
    \caption{ Intensity-weighted velocity map (moment 1) of the ground-state  HCN~$J=3\rm{-}2$ transition. 
    The contours range from 2850~\kms~ to 3250~\kms~ with steps of 50~\kms.
    The black line indicates the major axis of the two-dimensional Gaussian fit to the peak intensity of the continuum.
    The rough location and direction of the outflow (this work) is indicated by the dashed arrows.
    }
    \label{fig:HCN_mom1}
\end{figure}

Both the vibrational ground and first excited states of HCN $3-2$ have been detected towards \galaxy. 
Ground-state HCN $3-2$ emission is observed in a $\sim$0\farcs5 (87pc) structure, whereas the excited vibrational state (HCN-vib) is only detected in the innermost 80~mas (14~pc; Figure \ref{fig:Continuum}). 
Figures \ref{fig:HCN_mom0} and \ref{fig:HCN_mom1} show the respective integrated intensity and intensity-weighted velocity maps of the ground-state HCN~$J=3\rm{-}2$ line. 
The intensity-weighted velocity maps only include pixels $>3\times$~rms of the data cube. 
The ground-state line is generally seen in emission, though it can be seen in absorption towards the peak of the continuum (Figure \ref{fig:HCN_mom0}).

The ground-state HCN intensity-weighted velocity map indicates a two-sided outflow, with the blueshifted lobe proceeding to the northeast and the redshifted lobe proceeding to the southwest.  
The outflow  has observed velocities $>100$~\kms\ in both directions (Figure~\ref{fig:HCN_mom1}). 
The HCN-vib intensity-weighted velocity map (Figure~\ref{fig:HCN_vib_mom1}) shows rotation in the centremost 20~pc.

\subsection{Vibrational HCN morphology \label{sec:HCNvibMorph}}
The rotational transitions of vibrationally excited HCN occur in the $\nu_2$=1  state, which has an energy above the ground state of 1024 K.
A critical density of $\sim10^{11}~\rm{cm}^{-3}$ is necessary to collisionally excite these transitions \citep{Ziurys1986}, making HCN-vib emission a viable probe of the opaque nuclei of galaxies \citep[e.g.,][]{Martin2016}.
The high critical density likely means that radiative excitation populates the vibrationally excited states.
\citet{Aalto2015a} claim HCN-vib can directly trace the structure and dynamics of optically thick dust cores. 

\begin{figure}[]
    \centering
    \includegraphics[width=0.49\textwidth]{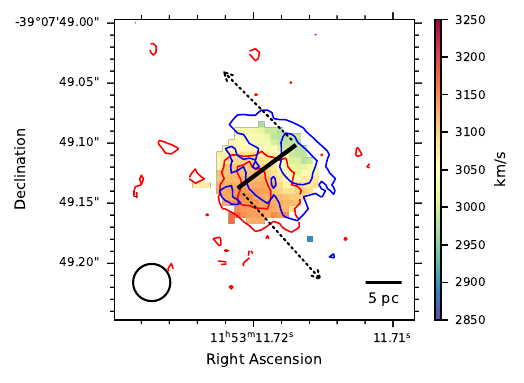}
    \caption{ Intensity-weighted velocity field (moment 1) of the vibrationally excited  HCN-vib transition, showing the base of the outflow.
    The black line indicates the major axis of the two-dimensional Gaussian fit to the peak intensity of the continuum.
    The rough location and direction of the outflow is indicated by the dashed arrows.
    Rotation is seen along the major axis of the continuum, with the blueshifted side to the northwest.
    The red and blue contours are integrated flux maps of $|v|>100$~\kms\ HCN-vib emission with levels at three and eight times 0.012~Jy~\kms . 
    These contours show that the HCN-vib emission is extended along the outflow, and that the outflow is launched from similarly rotating sides of the nucleus.
    }
    \label{fig:HCN_vib_mom1}
\end{figure}

The HCN-vib traces a compact region about the nucleus approximately 80~mas (14~pc) in diameter. 
The velocity field shows rotation with a projected velocity of roughly 100~\kms.
Integrated flux maps ($|v|>100$~\kms\ with respect to the systemic velocity) indicate that emission is extended along the axis of the molecular outflow (Figure \ref{fig:HCN_vib_mom1}).

\begin{figure*}[]
    \sidecaption
    \centering
    \includegraphics[width=0.7\textwidth]{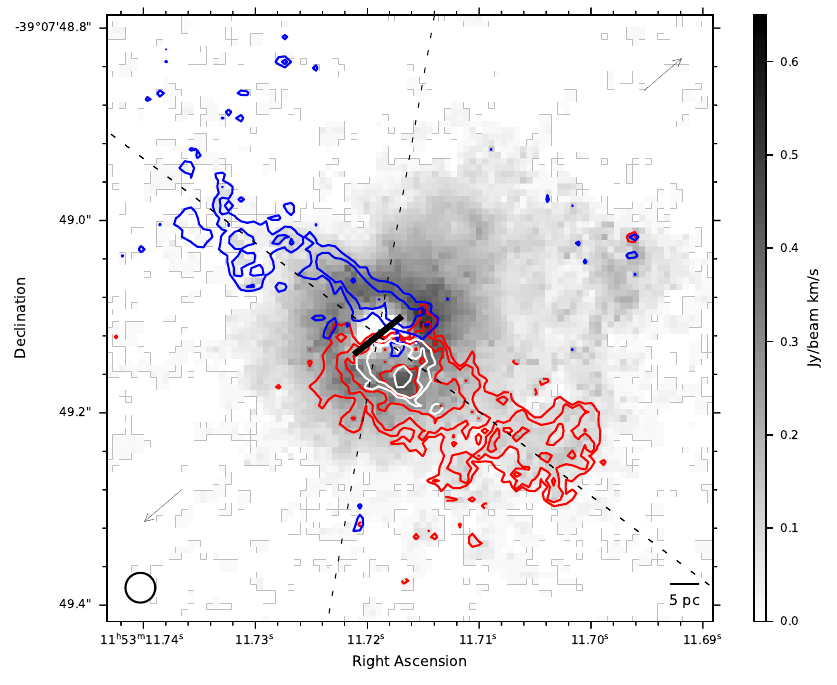}
    \caption{ Integrated ground-state HCN~$\nu=0$~$J=3\rm{-}2$ emission (gray scale), with  $|v|>100$~\kms\ (with respect to the systemic velocity) contours at 0.015~$\times$ (3, 5, 8, and 13) Jy~\kms. 
    The blue contours show the approaching outflow, and the red contours show the receding outflow.
    The white contours show the $v>200$~\kms\ gas.
    The black line indicates the major axis of the two-dimensional Gaussian fit to the peak intensity of the continuum.
    The dashed black line indicates the orientation of the outflow indicated by \citet{Pereira-Santaella2016,Pereira-Santaella2020}, and the arrows indicate the tentative direction of a radio elongation from \citet{Hekatelyne2020}.
    }
    \label{fig:high_vel_contours}
\end{figure*}

\section{Discussion \label{sec:discussion}}

\subsection{Outflow properties \label{sec:outflowmorph}}
\subsection{Outflow morphology}

\begin{figure*}[]
    \sidecaption
    \centering
    \includegraphics[width=0.7\textwidth]{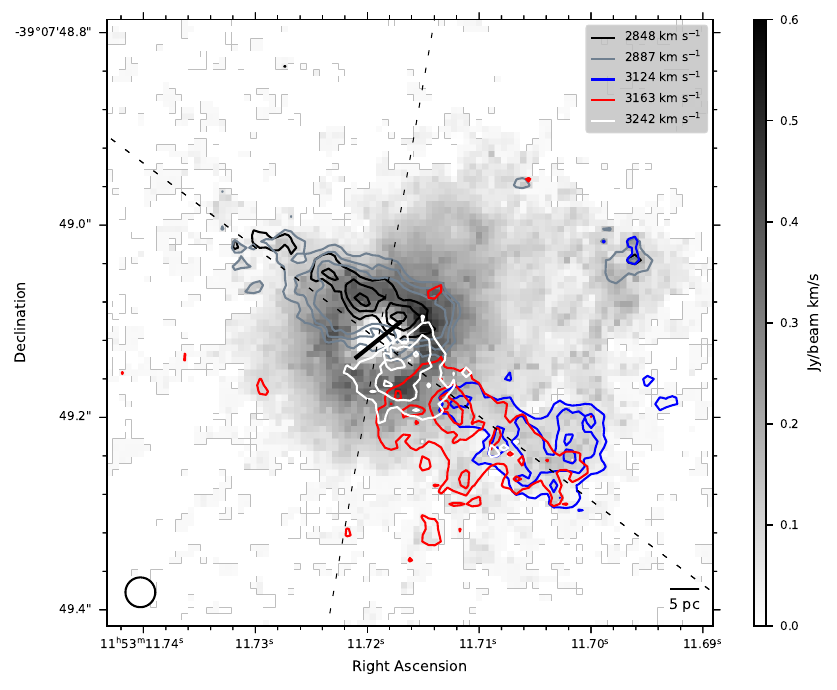}
    \caption{ Integrated ground-state HCN~$\nu=0$~$J=3\rm{-}2$ emission (gray scale), and 36~MHz ($\sim 40$~\kms) wide channel maps with contours at 0.14~$\times$ (5, 7, and 9) mJy.
    The blue contours show the approaching side of the redshifted lobe, and the red contours show the receding side of the redshifted lobe.
    The white contours show some of the highest velocity gas.
    Black and gray contours show the blueshifted lobe. 
    }
    \label{fig:renzo}
\end{figure*}

To image the outflow, we created two HCN 3-2 integrated flux maps.
We included channels that, by eye, are dominated by the outflow $|v-v_{sys}|>100$~\kms (Figure~\ref{fig:high_vel_contours}).
We observed the outflow out to 0\farcs3 (52~pc) in HCN 3-2. 
We also applied a Hanning kernel data cube, degrading the spectral resolution to 36~MHz ($\sim 40$~\kms) and thus improving the rms to 0.14 mJy beam$^{-1}$.
Figure \ref{fig:renzo} shows five individual channels of the ground-state HCN line. 

In Figure \ref{fig:high_vel_contours} the black bar shows the orientation of the nuclear region for reference, and the base of the outflow is seen separated by the approximate width of the continuum. 
There exists a velocity shift along the major axis of the continuum that is consistent with the rotation observed in the HCN-vib line.
The velocity shift in the outflow is approximately $\pm$150~km/s (projected). 
The gas is roughly consistent  with the 200~pc to kiloparsec scale outflow observed by \citet{Pereira-Santaella2016,Pereira-Santaella2020} with position angles between -10\deg \ and 53\deg \ (Figure \ref{fig:high_vel_contours}) and with velocities  of $|v|< 300$\kms. 

What is spectacular about the outflow morphology is that the launching regions are apparent and connected to the rotating nuclear structure in the innermost $\sim12$~pc.
The launch region of the blueshifted lobe is clearly separated from the redshifted lobe.
Additionally, in Figure \ref{fig:renzo} the blue contours (3124~\kms) indicate the approaching  side of the redshifted lobe and the red contours (3163~\kms) show the receding side of the redshifted lobe.
There is a clear east-west shift in the redshifted lobe, indicative of rotation in the outflow (see figure 12 from \citealt{aalto2020}). 
The 3242\kms\ contours (white) also show evidence of a highly inclined disk, suggesting the inclination derived from the continuum is accurate or a lower limit. 
The flux density drops from $\sim7\sigma$ to zero within one beam, forming a straight line consistent with an almost edge-on, or near edge-on, disk. 

To estimate the molecular mass of the outflow, we measured the integrated flux in the maps of the out flowing gas created in section \ref{sec:outflowmorph}.
We first converted the HCN $J$=3-2 line luminosity to HCN 1-0, assuming a line ratio of 0.5 as in \citet{Aalto2015a}. 
With this, we utilized the  L(HCN)-to-\mh\ conversion factor, calibrated for HCN J=1-0 \citep{Gao2004}, and modified for ULIRGs and HCN-bright environments by  \citet{Garcia-Burillo2012}. 
The conversion is $M_{\rm dense}\approx$ 3.1 \Msun/L$^{\prime-1}$ $\times \rm{L}(\rm{HCN}~1\rm{-}0)$, where the units on L$^{\prime}$ are K~\kms~pc$^{-2}$.
We measured a total HCN $J=3\rm{-}2$ luminosity  from the blue and red outflows of $3.4\pm0.1\times10^6$~K~\kms~pc$^{-2}$, and obtained an outflow mass of $2.1\times10^7$~\Msun~ within a radius of 52~pc.
The mass is a lower limit as only the higher velocity gas is included in the outflow maps, and portions of the blueshifted lobe are not included due to line-of-sight effects as discussed in section \ref{sec:powersource}.

\subsection{Outflow momentum}

We now explain how we estimated the mechanical luminosity of the outflow to attempt to understand the power source.
Momentum-driven outflows may exceed $L_{\rm Edd}/c$ up to a factor of five \citep{Roth2012}, whereas energy-driven outflows may exceed  $10 L_{\rm Edd}/c$ \citep{Costa2014}. 
Determining the mechanical luminosity of the molecular outflow is difficult due to substantial uncertainties on the molecular mass and inclination.
The kinetic energy of the outflow is $E_{\rm outflow}=(0.5  M_{\rm out} \times V_{\rm out}^2) + E_{\rm turb}$ \citep{Veilleux2001}.
We assumed that turbulent kinetic energy is insignificant. 
We adopted the inclination derived from the axial ratio of the continuum of 63.4$\pm$0.5\deg\ and the outflow mass derived in the previous paragraph.
For a projected velocity of 217$\pm$14~\kms, taken from the average of the maximum observed velocities of the red (3322~\kms) and blue (2828~\kms) lobes and one channel width uncertainty, the outflow velocity is $V_{\rm out}=515\pm33$~\kms.
The total kinetic energy of the outflow is then $(6.5\pm0.8)\times10^{55}$~erg.
For a radial extent of the outflow of 52~pc, measured from the extent of the 3.3 Jy~\kms\ ground-state HCN contour, this kinetic energy results in a mechanical luminosity of $(5.8\pm1.1)\times10^9$~\Lsun.
It is important to note that the following result is a lower limit to the outflow rate and momentum flux. 
The outflow rate and momentum flux are respectively $231\pm15$~\Msun~yr$^{-1}$ and $(6.2\pm0.8)\times10^{12}$~\Lsun$/c$.

For \galaxy, \citet{Gonzalez-Alfonso2021} estimate the mass of the black hole, from the $M_{\rm BH}/\sigma$ relation, to be between 5$\times10^6$~\Msun\ and 8$\times10^6$~\Msun.
The corresponding Eddington luminosity is 2.1$\pm0.5\times10^{11}$~\Lsun.
The momentum flux of the outflow is $30\pm7 \times L_{\rm Edd}/c$ for a $6.5\times10^6$~\Msun\ black hole.
It is unlikely that the black hole is much larger than this, as the enclosed mass estimated from the HCN-vib emission within 70~mas (12~pc) and a rotation of 100~\kms\ is 7$\times10^7$~\Msun. 
Furthermore, the outflow momentum flux exceeds the total infrared luminosity of the galaxy ($L_{\rm IR}/c$) by a factor of $63\pm8$.

The analysis of the momentum flux reveals that the outflow is unlikely to be momentum driven by an AGN, as this exceeds the factor of five limit estimated by \citet{Roth2012}. 
It is similarly unlikely to be star-formation driven, since starburst-driven outflows appear to have a momentum limit of 1-3 $L_{\rm bol}$/c \citep{Geach2014}. 
In LIRGs, $L_{\rm bol} \approx L_{\rm IR}$ \citep{Sanders1996}.
Hence, even if all the star formation luminosity was emerging from the inner launch region of the outflow, the momentum flux is too high by a factor of ten. 
In addition, we know that only a small fraction of the star formation is contained within the nucleus, as revealed by Pa$\alpha$ maps \citep{Sanchez-Garcia2022}, so the infrared momentum flux of the entire galaxy acts as a conservative upper limit.
The remaining possibilities are an energy-driven AGN wind \citep[e.g.,][]{Faucher-Giguere2012,Costa2014}, which is capable of powering momentum fluxes in excess of $10 \times L_{\rm Edd}/c$, a radio jet, or a MHD-powered wind.

\subsection{The outflow power source \label{sec:powersource}}

It is difficult to ascertain which of the remaining mechanisms, from the previous section, dominate in \galaxy. 
A radio-jet-powered outflow would technically be possible.
\galaxy\ shows tentative evidence for a radio jet (Figure 3 from \citealt{Hekatelyne2020}); however, it is incorrectly oriented to power the molecular outflow (Figure \ref{fig:high_vel_contours}).
The radio extension is also not detected in the higher-resolution observations by \citet{Baan2006}.

A central constituent of an energy-driven AGN wind is a hot ionized component \citep{Faucher-Giguere2012}. 
The hot AGN wind entrains molecular gas in the torus, and the torus collimates the outflow \citep[e.g.,][]{Garcia-burillo2019}.
\citet{Cazzoli2014} suggest that an ionized outflow is sometimes seen in the northeast of the \galaxy. 
The detection is very tentative, and claimed to trace the outer edge of the wind.
However, analysis by \citet{Bellocchi2013} and \citet{Pereira-Santaella2016} do not confirm the existence of an ionized outflow.
The maximum ionized outflow velocity is estimated to be $\sim100$\kms\ \citep{Arribas2014}.
In addition, the X-ray,  mid-infrared,  optical, and radio spectra do not corroborate the existence of an AGN. 
\galaxy's X-ray spectrum neither shows the Fe~K$\alpha$ line indicative of Compton-thick AGN \citep[][]{Pereira-Santaella2011}, nor does the mid-infrared spectrum show the high ionization AGN lines [{{Ne \sc {\scriptsize V}}}] 14.32~$\mu$m, [{{Ne \sc {\scriptsize V}}}] 24.32~$\mu$m, or [{{O \sc {\scriptsize IV}}}] 25.89~$\mu$m \citep[][]{Pereira-Santaella2010b}.
{{[N\sc {\scriptsize II}]}}/H$\alpha$ and {{[O\sc {\scriptsize III}]}}/H$\beta$ diagnostics also corroborate the lack of an AGN.
\galaxy\ has a {{H\sc {\scriptsize II}}} type spectrum \citep{VandenBroek1991,Pereira-Santaella2011}, and it is missing the corresponding radio diagnostics of an AGN \citep{Baan2006}. 
These are crucial missing components to an energy-driven AGN outflow.

The key lies in the morpho-kinematic structure. 
Figures \ref{fig:high_vel_contours} and \ref{fig:renzo} clearly reveal the imprint of nuclear rotation in the outflow. 
The redshifted and blueshifted lobes of the outflow appear to be launched from the respective rotating sides of the inner nuclear structure.
The redshifted lobe also shows an east-west velocity shift (Figure \ref{fig:renzo}).
There also appears to be a deficiency of material on the blueshifted lobe of the outflow compared with the redshifted lobe and higher velocity gas near the nucleus.
Furthermore, the outflow achieves a momentum flux exceeding $30~L_{\rm Edd}/c$ within a radius of 100~pc.
A 10$^8$~\Msun\ black hole accreting at the Eddington accretion rate can produce a momentum flux of $\sim26~L_{\rm Edd}/c$, but this requires kiloparsec scales for the outflow to cool \citep{Costa2014}.
As \galaxy's outflow significantly exceeds the momentum flux upper limit for dusty radiation-driven outflows on $10\sim100$~pc scales, and since the SMBH mass cannot exceed the mass enclosed of 7$\times10^7$~\Msun, it is highly likely the dominant mechanism driving the outflow is not an energy-driven AGN. 
In the protostellar analogy proposed by \citet{Gorski2023}, this makes sense in the context of a rotating infalling envelope and a MHD-powered outflow \citep[e.g.,][]{Sakai2014,Oya2014,Oya2021}.
The lack of material is a line-of-sight effect where the redshifted side of the blueshifted lobe is confused with, or obscured by, the gas rotating closer to the systemic velocity of the galaxy.
The outflow reappears once it emerges from the apparent extent of the envelope. 
The rotation of outflows is a strong indication of magnetic acceleration \citep[][\& references therein]{Bjerkeli2016, Proga2007}. 
\citet{Pudritz2005} concluded that jets and outflows are inherent to the collapse of magnetized cores, and their properties are independent of the system's mass.
Here, the outflow emerges from the centre of a rotating disk with a significant amount of angular momentum indicative of a MHD wind.

\section{Conclusion}\label{sec:conclusions}

The astronomical community actively debates whether galactic outflows are driven by jets, mechanical winds, or radiation.
In this Letter, we present compelling evidence that the outflow in \galaxy\ is powered by a different mechanism, a MHD wind launched prior to the ignition of an AGN. 
The overall morpho-kinematic structure appears to be consistent with a rotating disk and a rotating outflow. 
Analysis of the momentum flux excludes mechanically driven AGN, energy-driven AGN, or starburst-driven winds.
The lack of an ionized outflow, or corresponding X-ray, radio, infrared, or optical counterparts suggests that there is no AGN or very weak AGN incapable of powering the outflow.
What remains is a MHD wind, as evidenced by the rotating outflow.
These results paint the following picture: that the nuclear evolution of galaxies and the growth of SMBHs is analogous to the growth of hot cores or protostars; feedback may not be necessary to drive a galactic wind; and nuclear evolution is regulated, at least in part, by MHD processes. 

\begin{acknowledgements}

S.A., S.K., K.O., C.Y. gratefully acknowledge funding from the European Research Council (ERC) under the European Union's Horizon 2020 research and innovation programme (grant agreement No. 789410).
S.V acknowledges support from the European Research Council (ERC) under the European Union’s Horizon 2020 research and innovation program MOPPEX 833460.

M.G. acknowledges support from the Nordic ALMA Regional Center (ARC) node based at Onsala Space Observatory. 
The Nordic ARC node is funded through Swedish Research Council grant No 2017-00648.

EG-A thanks the Spanish Ministerio de Econom\'{\i}a y Competitividad for support under project PID2019-105552RB-C41.

This paper makes use of the following ALMA data: ADS/JAO.ALMA\#2019.1.01612.S. ALMA is a partnership of ESO (representing its member states), NSF (USA) and NINS (Japan), together with NRC (Canada), MOST and ASIAA (Taiwan), and KASI (Republic of Korea), in cooperation with the Republic of Chile. The Joint ALMA Observatory is operated by ESO, AUI/NRAO and NAOJ.

\end{acknowledgements}
\bibliographystyle{aa} 
\bibliography{BB_COMs,BB_ConQuest,BB_eso320,BB_misc} 

\begin{appendix}

\section{Supplemental plots and figures}

\begin{figure}[]
    \centering
    \includegraphics[width=0.5\textwidth]{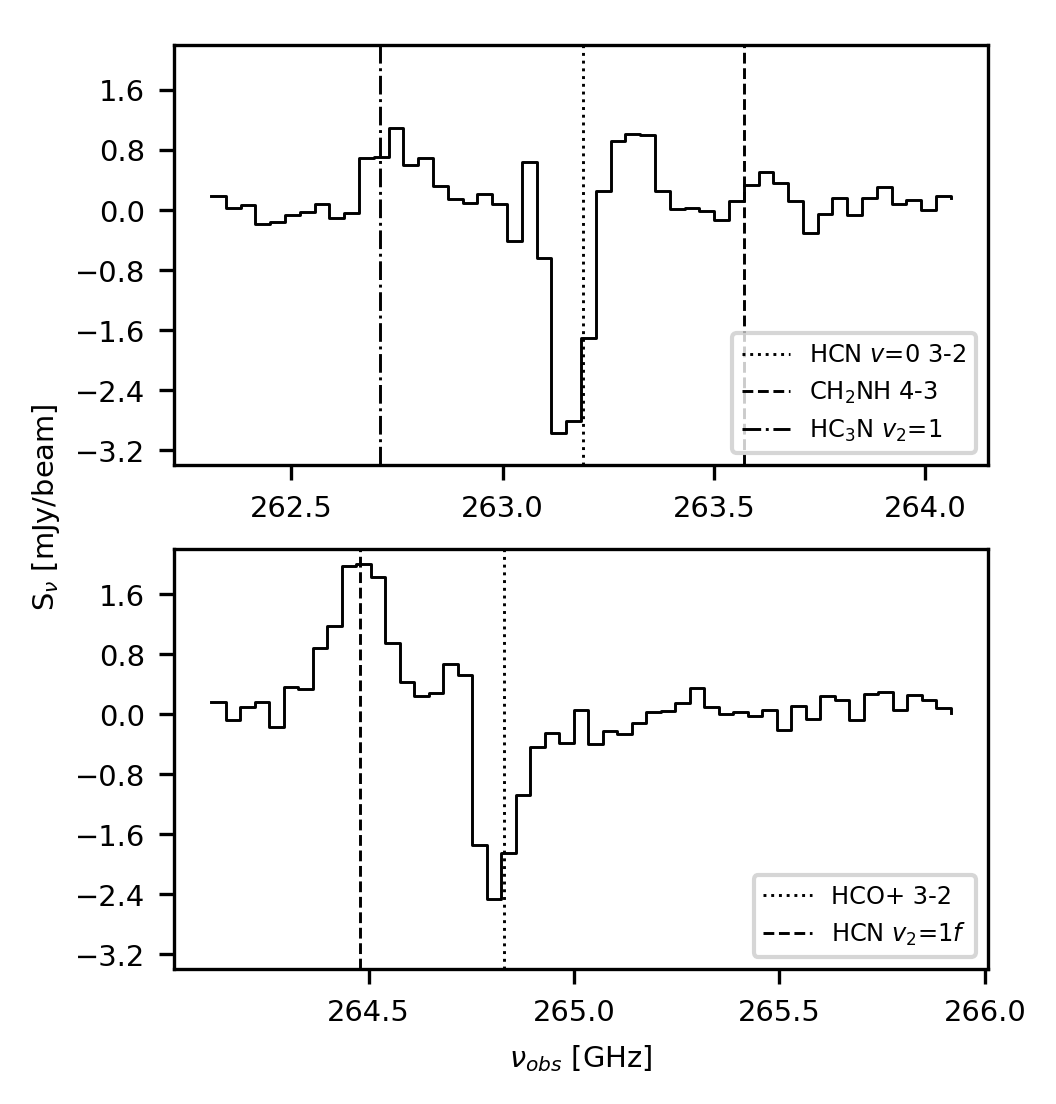}
    \caption{Spectrum through the peak continuum pixel of two 1.875 GHz wide bands centred at 265.005 GHz and 263.182 GHz with 32~GHz ($\sim$40~\kms) channels. Spectral lines of particular interest are labelled with vertical lines.}
    \label{fig:spectrum}
\end{figure}

\begin{figure*}[]
    \centering
    \includegraphics[width=1.0\textwidth]{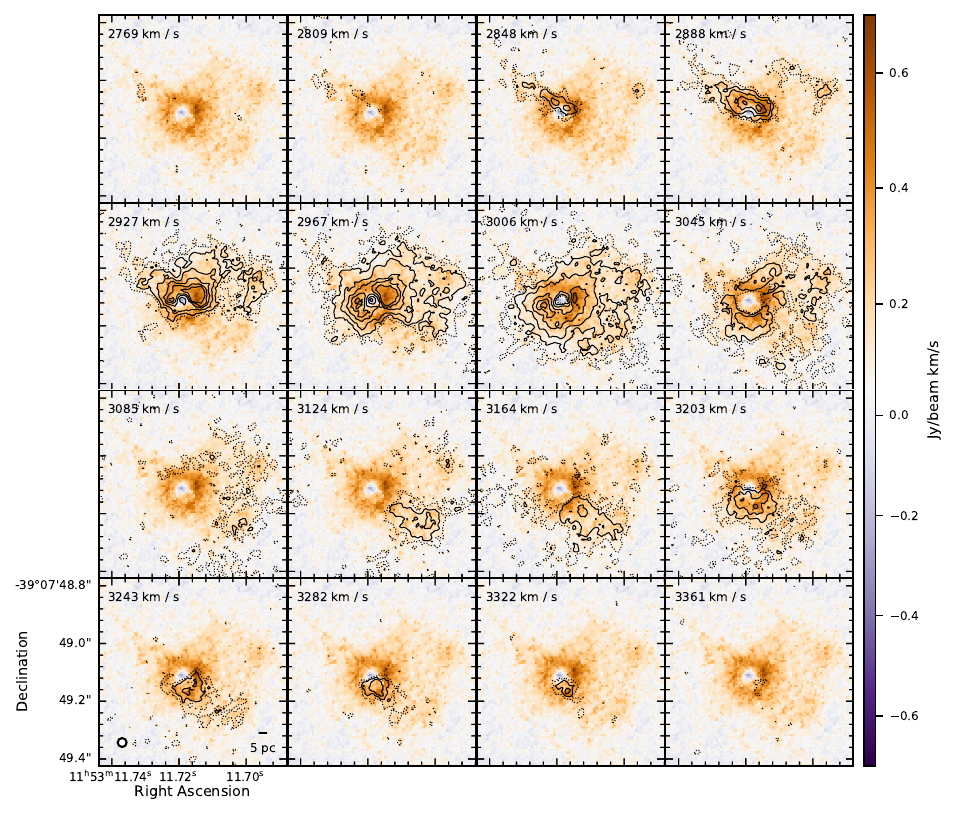}
    \caption{ Integrated ground-state HCN~$\nu=0$~$J=3\rm{-}2$ emission, and 36~MHz ($\sim 40$~\kms) wide channel maps with solid contours at 0.14~mJy~$\times$ (5, 8, 12,15, and 18). The lowest 0.14~mJy~$\times$~3 contour is dotted.
    }
    \label{fig:channels}
\end{figure*}

\begin{figure*}[]
    \centering
    \includegraphics[width=1.0\textwidth]{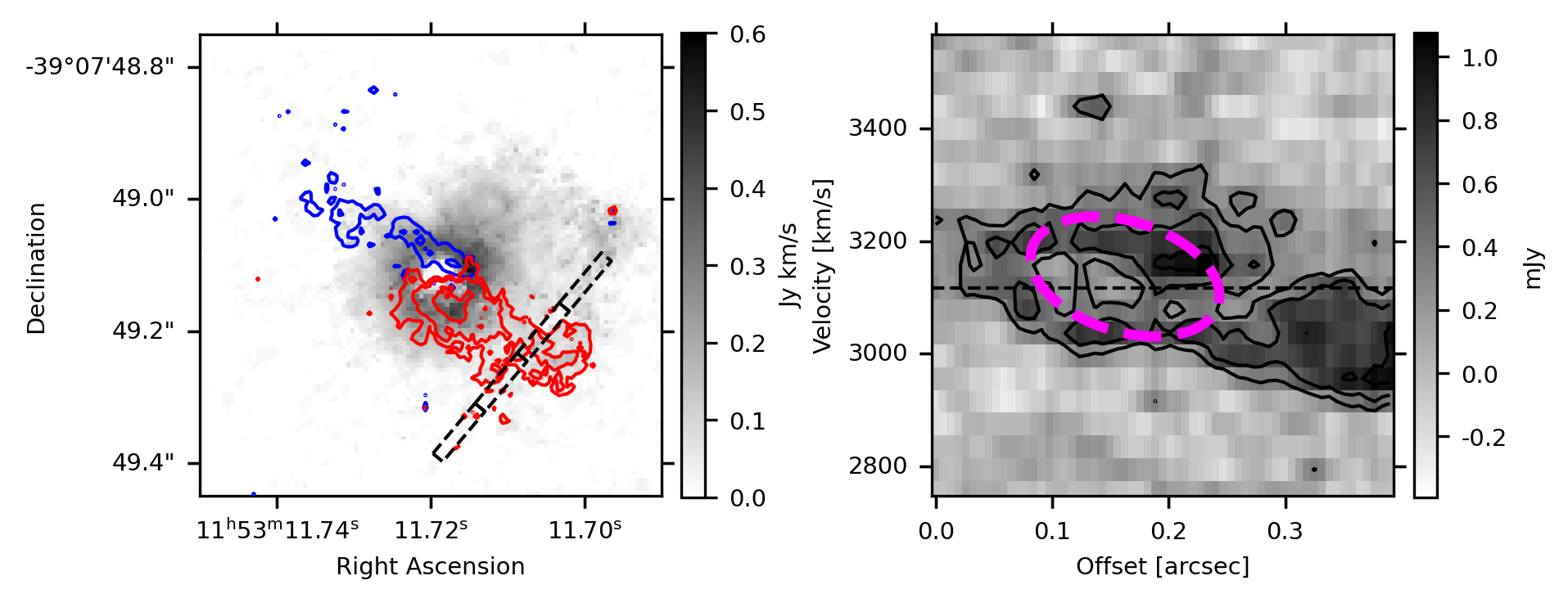}
    \caption{[Left] Integrated ground-state HCN~$\nu=0$~$J=3\rm{-}2$ emission (gray scale), with  $|v|>100$~\kms\ (with respect to the systemic velocity) contours at 0.015~$\times$ (3, 5, 8, and 13) Jy~\kms. 
    The blue contours show the approaching outflow, and the red contours show the receding outflow.
    The dashed black bar indicates the region of the position velocity cut through the redshifted outflow lobe. 
    [Right] Position velocity cut with 0.11~mJy~$\times$ (3, 5, and 8) contours. 
    The dashed magenta cartoon ellipse shows the characteristic ring of emission indicative of a rotating outflow.
    The dashed black line shows the systemic velocity of \galaxy.
    }
    \label{fig:rlobepv}
\end{figure*}

\begin{figure*}[]
    \centering
    \includegraphics[width=1.0\textwidth]{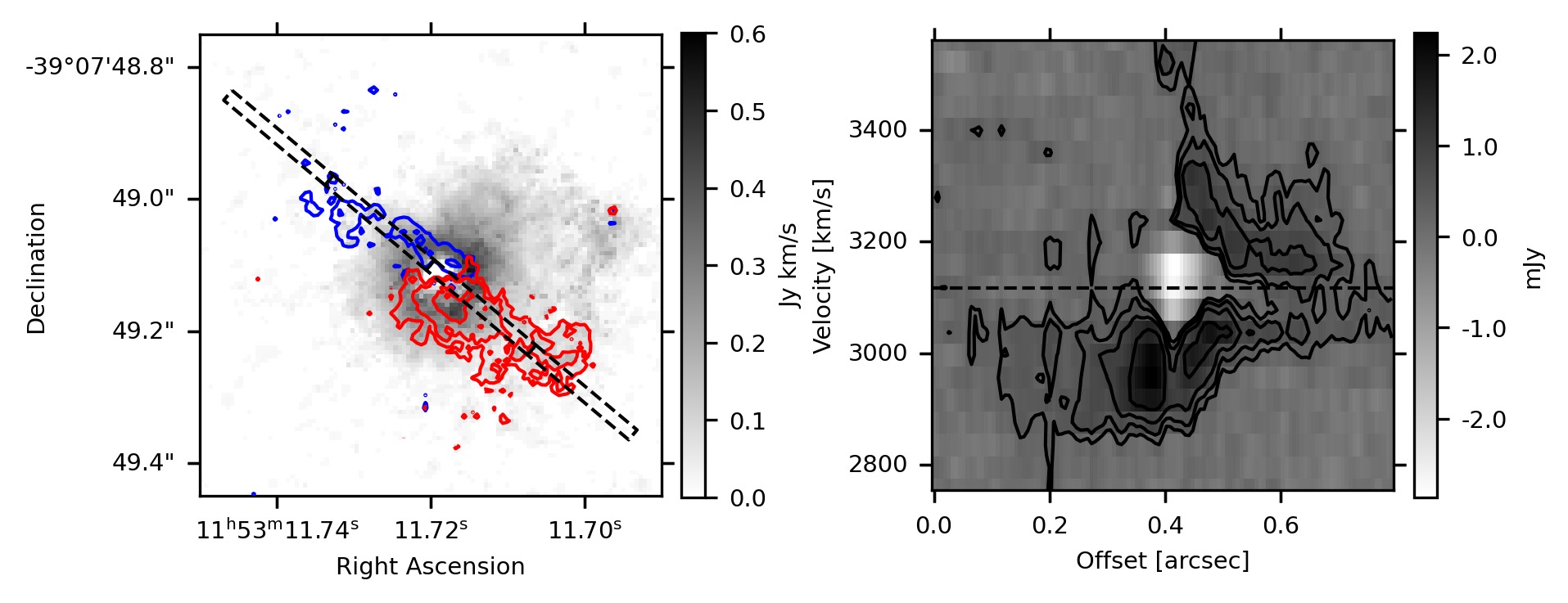}
    \caption{ [Left] Integrated ground-state HCN~$\nu=0$~$J=3\rm{-}2$ emission (gray scale), with  $|v|>100$~\kms\ (with respect to the systemic velocity) contours at 0.015~$\times$ (3, 5, 8, and 13) Jy~\kms. 
    The blue contours show the approaching outflow, and the red contours show the receding outflow.
    The dashed black bar indicates the region of the position velocity cut through the redshifted outflow lobe. 
    [Right] Position velocity cut with 0.11~mJy~$\times$ (3, 5, and 8) contours. 
    The dashed black line shows the systemic velocity of \galaxy.
    }
    \label{fig:outflowpv}
\end{figure*}

\end{appendix}

\end{document}